# Nitrogen isotopic fractionation during abiotic synthesis of organic solid particles

*Maïa Kuga[1*], Nathalie Carrasco[2,3], Bernard Marty[1], Yves Marrocchi[1], Sylvain Bernard[4], Thomas Rigaudier[1], Benjamin Fleury[2] and Laurent Tissandier[1]*

[1] CRPG-CNRS, Université de Lorraine, 15 rue Notre Dame des Pauvres. 54500 Vandoeuvre-les-Nancy, France

[2] Université de Versailles St Quentin, LATMOS, Université Paris 6 Pierre et Marie Curie, CNRS. 11 bvd d'Alembert, 78280 Guyancourt, France

[3] Institut Universitaire de France, 103 Boulevard Saint-Michel, 75005 Paris, France

[4] IMPMC, MNHM, UPMC, UMR CNRS 7590, 61 rue Buffon, 75005 Paris, France

*corresponding author.
E-mail address : mkuga@crpg.cnrs-nancy.fr

Keywords : nitrogen isotopesorganics, plasma, Solar System, Titan's atmosphere, early Earth

**Abstract**

The formation of organic compounds is generally assumed to result from abiotic processes in the Solar System, with the exception of biogenic organics on Earth. Nitrogen-bearing organics are of particular interest, notably for prebiotic perspectives but also for overall comprehension of organic formation in the young solar system and in planetary atmospheres. We have investigated abiotic synthesis of organics upon plasma discharge, with special attention to N isotope fractionation. Organic aerosols were synthesized from $N_2$-$CH_4$ and $N_2$-CO gaseous mixtures using low-pressure plasma discharge experiments, aimed at simulating chemistry occurring in Titan's atmosphere and in the protosolar nebula, respectively. The nitrogen content, the N speciation and the N isotopic composition were analyzed in the resulting organic aerosols. Nitrogen is efficiently incorporated into the synthesized solids, independently of the oxidation degree, of the $N_2$ content of the starting gas mixture, and of the nitrogen speciation in the aerosols. The aerosols are depleted in $^{15}N$ by 15-25 ‰ relative to the initial $N_2$ gas, whatever the experimental setup is. Such an isotopic fractionation is attributed to mass-dependent kinetic effect(s).

Nitrogen isotope fractionation upon electric discharge cannot account for the large N isotope variations observed among solar system objects and reservoirs. Extreme N isotope signatures in the solar system are more likely the result of self-shielding during $N_2$ photodissociation, exotic effect during photodissociation of $N_2$ and/or low temperature ion-molecule isotope exchange. Kinetic N isotope fractionation may play a significant role in the Titan's atmosphere. In the Titan's night side, $^{15}N$-depletion resulting from electron driven reactions may counterbalance photo-induced $^{15}N$ enrichments occurring on the day's side. We also suggest that the low $\delta^{15}N$ values of Archaean organic matter (Beaumont and Robert, 1999) are partly the result of abiotic synthesis of organics that occurred at that time, and that the subsequent development of the biosphere resulted in shifts of $\delta^{15}N$ towards higher values.

# 1 Introduction

Organic compounds are widely distributed in the Solar System, from very simple to more complex molecules. In contrast to what happen on Earth where organic matter has been mostly synthesized by life since the Archaean, and possibly, the Hadean, eon(s), the formation of organic compounds is generally assumed to result from abiotic processes in the Solar System. Extraterrestrial organics are found mostly as amino-acids and kerogen-like material in primitive meteorites and as organic haze in planetary atmospheres such as Titan's. Among such organic molecules, the nitrogen-bearing ones are of particular prebiotic interest, as nitrogen is a key element of proteins and nucleic acids. Furthermore, nitrogen seems to play a pivotal role during the production of organic aerosols in the $N_2$-rich atmosphere of Titan (Israël et al., 2005, Carrasco et al., 2013).

Nitrogen has two stable isotopes, $^{14}N$ and $^{15}N$. The $^{15}N/^{14}N$ ratio (3.676 x $10^{-3}$ for the terrestrial atmospheric $N_2$) is often expressed in permil deviation to terrestrial atmospheric $N_2$ isotope composition (AIR) as $\delta^{15}N = [(^{15}N/^{14}N)_{sample} / (^{15}N/^{14}N)_{AIR} -1] \times 1000$ (in ‰). Relative abundances of $^{14}N$ and $^{15}N$ fractionate upon physical, chemical and biological transformations of N-bearing compounds. Remarkably, the isotopic composition of nitrogen presents dramatic variations among solar system objects and reservoirs, which are not fully understood. $N_2$ was probably the main N-bearing species in the protosolar nebula (PSN, Grossman, 1972) and was $^{15}N$-poor ($^{15}N/^{14}N$ = 2.27 ± 0.03 x $10^{-3}$, that is, $\delta^{15}N$ = -383 ± 8 ‰; Marty et al., 2011) whereas all other objects and reservoirs of the Solar System (with the exception of a few like Jupiter's atmosphere) are richer in $^{15}N$ by several hundreds to thousands of permil. Most meteorite families and the inner planetary bodies including the Earth have comparable $\delta^{15}N$ values within a few tens of ‰ whereas cometary CN and HCN are enriched in $^{15}$N by a factor of 3 relative to the PSN value (Bockelée-Morvan et al., 2008). $^{15}$N enrichments can be dramatic at the micron scale in meteoritic organics, with $\delta^{15}$N values up to 5,000 ‰ (Briani et al., 2009). Some of these enrichments may be related to atmospheric processing, as proposed for the atmosphere of Mars ($\delta^{15}$N$_{N2}$ = +660 ‰, Owen et al., 1977), but in other cases these $^{15}$N enrichments relative to the PSN nitrogen require other types of extensive isotope fractionation that are poorly understood.

Exothermic ion-molecule reactions at low temperature might have led to extensive N isotope fractionation under specific cold yet dense environments (e.g., dense cores, outer solar system, Terzieva and Herbst, 2000, Rodgers and Charnley, 2008, Aleon, 2010, Hily-Blant et al., 2013). Alternatively, photodissociation of $N_2$ associated or not to self-shielding (Clayton, 2002, Lyons et al., 2009, Chakraborty et al., 2013), might have led to $^{15}$N-rich radicals prone to incorporation into forming organics. Such a photochemical induced fractionation has been invoked to address the $^{15}$N-rich HCN relative to $N_2$ in Titan's atmosphere (Vinatier et al., 2007, Liang et al., 2007, Croteau et al., 2011). However, those theoretical studies have focused on the isotopic composition of very simple gaseous N-bearing compounds, and the propagation and conservation of such a large N isotopic fractionation upon polymerization of organic solids has not yet been fully investigated.

Several experimental works have simulated the synthesis of gaseous and solid organic compounds in gas mixtures similar to Titan's atmosphere (see Coll et al., 2013, for a review) early Earth's atmosphere (Miller, 1953, Chang et al., 1983) or the PSN (Dzizckniec and Lumkin, 1981, Kerridge et al., 1989), either by UV photons or by electron energy deposition. However, only a few of them have focused on the nitrogen incorporation into refractory organics from $N_2$ dissociation (Trainer et al., 2012, Gautier, 2013), and the extent of related N isotope fractionation is essentially undocumented.

Whatever the original mechanism of isotope selection, $N_2$ dissociation and nitrogen compound ionization are believed to play an important yet not fully understood role. Plasmas were used in the present study because of the strong covalent bond of $N_2$, which needs energies above 9.8 eV (< 120 nm) to break. Recent experimental simulations have used VUV photons as incident energy but production of aerosols was not reported (Imanaka and Smith, 2010, Peng et al., 2013). So far, electron energy deposition stays the easiest energy source to simulate aerosol productions from irradiated gas mixtures. Here, we investigate the isotopic fractionation of nitrogen during synthesis of solid organics by plasma discharge using $N_2$-$CH_4$ and $N_2$-CO gaseous mixtures as proxies of Titan's atmosphere and the PSN, respectively.

## 2 Experimental methods

### 2.1 Aerosols production setups

Two experimental plasma setups were used for this study in order to produce nitrogen-rich aerosols: (i) the PAMPRE experiment (LATMOS, Guyancourt, France), designed to investigate Titan's ionosphere processes; and (ii) the Nebulotron experiment (CRPG, Nancy, France), dedicated to simulate young solar nebula processes. Main experimental conditions and plasma characteristics are described below and in the Table 1.

|  | **PAMPRE** | **Nebulotron** |
|---|---|---|
| gas mixture | $N_2$ (90-99%) - $CH_4$ | $N_2$ (20%) - CO |
| type of electric discharge | RF (13,56 MHz) | microwave (2.45 GHz) |
| size of the reactor (height * diameter) | 40*30 cm | 10*0.8 cm |
| volume of the plasma ($cm^3$) | 740 | 2.5 |
| injected power (W) | 30 | 30 |
| $W/cm^3$ | 0.04 | 12 |
| pressure (mbar) | 0.9 | ~ 1 |
| Neutral temperature | 300 - 350 K [a] | ≥ 1000 K (estimated[c,d]) |
| Electron density ($cm^{-3}$) | $2 \cdot 10^8$ [a] | $5 \cdot 10^{10}$ - $1 \cdot 10^{11}$ (estimated[d]) |
| Average electron energy | 2 eV [b] | 2 eV (estimated[d]) |

[a] *Alcouffe et al. (2010)*  [c] *Es-Sebbar et al. (2009)*
[b] *Alves et al. (2012)*   [d] *Gries et al. (2009)*

**Table 1**: Qualitative comparison of the two plasma setups used in this study.

### 2.1.1 The PAMPRE experiment

The PAMPRE (for *Production d'Aérosols en Microgravité par Plasma REactif*) experiment consists of a stainless steel reaction chamber, where a radiofrequency discharge (RF, 13.56 MHz) is generated between two electrodes in a metallic cage confining the plasma (Sciamma-O'Brien et al., 2010; Szopa et al., 2006, Fig. 1a).

A gas mixture of high purity $N_2$ and $CH_4$ is flowed continuously through the plasma discharge in which electrons dissociate and ionize $N_2$ and $CH_4$. This initiates chemical reactions and molecular growth that results in the production of hydrocarbons and N-bearing molecules that eventually end up forming solid particles. These solid particles grow up in suspension in the plasma and fall in a glass vessel surrounding the metallic cage. After typical runs of 8 h, the produced solid particles, orange to brown in color, are collected for ex-situ analysis.

In this work, experiments were performed with a continuous 55 sccm (standard cubic centimeter per minute) $N_2$-$CH_4$ gas flow, containing 1%, 2%, 5% and 10% $CH_4$ in $N_2$. These gas proportions are representative of the composition of Titan's atmosphere (Niemann et al., 2005, Waite et al., 2005). For all experimental conditions, the injected RF power was fixed to 30 W, the pressure in the reactor was 0.9 mbar and the neutral gas temperature, measured by Alcouffe et al. (2010), ranged from 310 K to 340 K depending on experimental conditions (Tables 1 and 2).

*2.1.2 The Nebulotron experiment*

The second experimental setup used in this study, the Nebulotron (CRPG-CNRS), consists of a vacuum glass line in which adjustable gas mixtures can be flown through a microwave (2.45 GHz) plasma discharge (Robert et al., 2011). The aim of this setup is to simulate processes occurring in a CO-$N_2$ atmosphere. CO and $N_2$ are believed to have been the main gaseous species hosting C and N in the protosolar nebula (Grossman, 1972). The experimental setup consists of a quartz reactor (Fig. 1b) where a gas mixture of CO (purity 99.5%) and $N_2$ (purity 99.995%) is flowed continuously through the plasma discharge at a pressure of 1 mbar. As for the PAMPRE experiment, electrons dissociate and ionize $N_2$ and CO. In this setup however, organic aerosols grow up on the quartz tube surfaces that are cooled down by compressed air. After a typical 6 hours-long experiment, the plasma is turned off and the glass line is pumped out for 12 hours before opening the reactor to atmospheric pressure. The orange to dark solids are then recovered by gently scratching the quartz tube and stored in microvials for ex-situ analysis. As for the PAMPRE setup, not all the solids produced can be collected and the mass production rates presented here (Table 2) must be considered as minimum values. No leak has been detected on the Nebulotron reactor when isolated from the pumping group and from the gas injection system. Nonetheless, contrary to the PAMPRE setup, this one is not fitted with a secondary turbo pump, allowing a $10^{-2}$ mbar limit vacuum only. This does not allow for sufficient desorption of water adsorbed on the reactor walls. This expected water contamination is discussed in the 3.1 Section. So far, the electronic features of the Nebulotron plasma setup have not been characterized yet but, as the volume of the plasma is much smaller than for the PAMPRE experiment, with a similar input power (30 W, Table 1), the temperature of neutral gases and the electron density in the Nebulotron setup are expected to be higher than in the PAMPRE plasma (Fridman, 2008, Es-Sebbar et al., 2009, Gries et al., 2009).

Here, a 6 hour-long experiment was performed with a 30 W injected power and a 6 sccm $CO-N_2$ gas mixture containing 80% CO in $N_2$ (Table 2). This experimental C/N ratio is comparable to what is estimated for the protosolar nebula (Lodders, 2003).

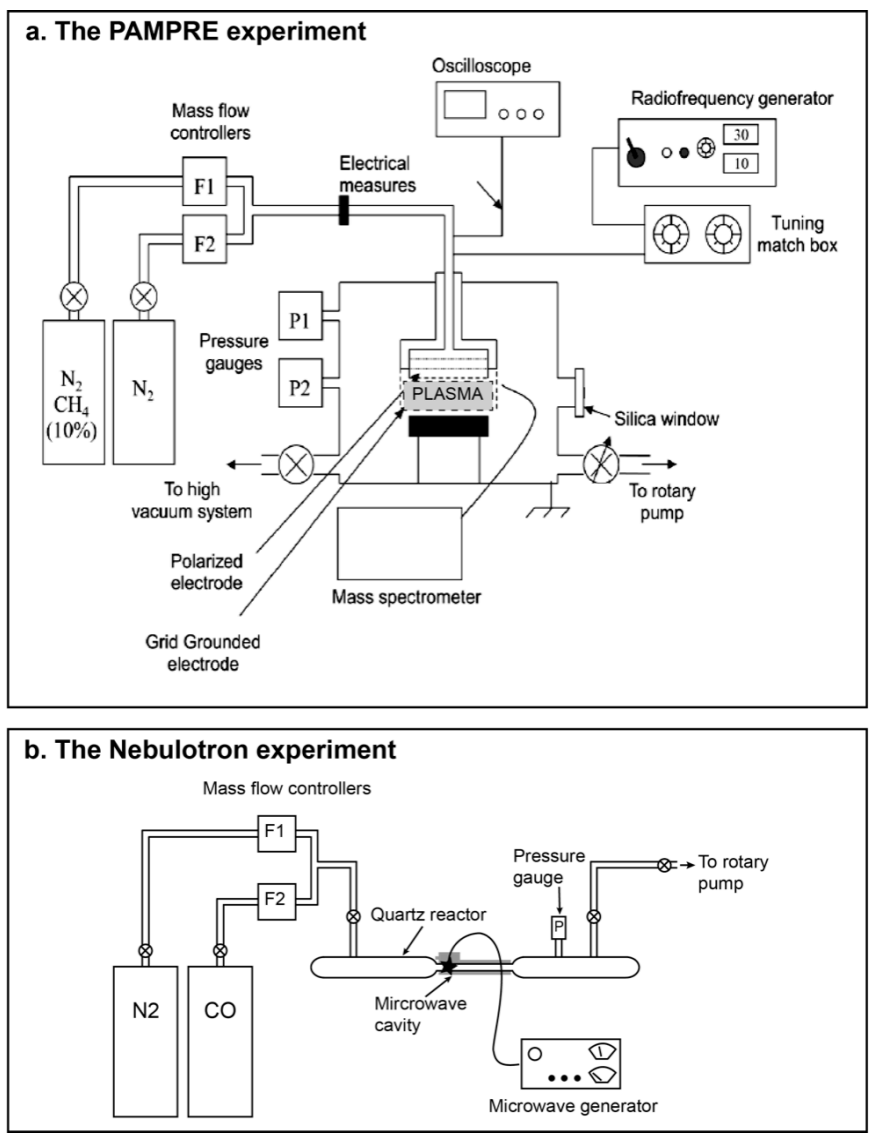

**Fig. 1**: Schematics of the PAMPRE (a) and the Nebulotron (b) experimental setups.

| a. The PAMPRE experiment ($N_2$-$CH_4$) | | | | | |
|---|---|---|---|---|---|
| %$CH_4$ i | %$CH_4$ ss[a] | Gas flow rate (sccm) | Power input (W) | Pressure (mbar) | aerosols production rate (mg.h$^{-1}$)[a] |
| 1 | 0.18± 0.01 | 55 | 30 | 0.9 | 18 |
| 2 | 0.42± 0.01 | 55 | 30 | 0.9 | 32 |
| 5 | 1.80± 0.03 | 55 | 30 | 0.9 | 33 |
| 10 | 5.53± 0.05 | 55 | 30 | 0.9 | 4 |
| [a]Sciamma O'Brien et al. (2010) | | | | | |
| b. The Nebulotron experiment ($N_2$-CO) | | | | | |
| %CO i | %CO ss | Gas flow rate (sccm) | Power input (W) | Pressure (mbar) | aerosols production rate (mg.h$^{-1}$) |
| 80 | nm | 6 | 30 | 1 | 6 |

*nm = not measured, i = initial, ss = at steady state, sccm = standard $c^3$/min*

**Table 2**: Experimental conditions of aerosol production in the PAMPRE and Nebulotron setups.

## 2.2 Elemental composition analysis of the aerosols

The elemental composition (C, H, N and O) of the Nebulotron aerosols has been analyzed at the Service d'Analyses Elementaires (Université de Lorraine, France) by combustion and pyrolysis techniques. The PAMPRE aerosol elemental composition was measured and reported previously by Sciamma-O'Brien et al., 2010.

## 2.3 Carbon- and Nitrogen-edge XANES analysis

Measurements were done using the STXM (*Scanning Transmission X-Ray Microscopy*) located on beamline 5.3.2.2 (Polymer STXM beamline - Kilcoyne et al., 2003) at the Advanced Light Source (ALS, Berkeley, USA). This beamline uses soft X-rays (250 - 600 eV) generated via a bending magnet while the electron current in the storage ring is held constant in top-off mode at 500 mA at a storage ring energy of 1.9 GeV. The microscope chamber was evacuated to 0.1 mbar after sample insertion and back-filled with helium. The energy calibration was carried out using the well-resolved 3p Rydberg peak at 294.96 eV of gaseous $CO_2$ for the C K-edge. In order to obtain partly X-ray transparent samples which do not completely absorb the incident light at the C and N K-edges, the PAMPRE and Nebulotron aerosols have been finely ground and deposited on SiN windows. The C- and N-XANES data shown here have been collected following the procedures for X-ray microscopy studies of radiation sensitive samples recommended by Wang et al., 2009. Alignment of images of stacks and extraction of XANES spectra were done using the aXis2000 software (ver2.1n). Normalization and determination of spectral peak positions were determined using the Athena software package (Ravel and Newville, 2005). Extensive databases of reference C- and N-XANES spectra are available for organic compounds (e.g., Leinweber et al., 2007; Solomon et al., 2009).

**2.4 Nitrogen isotope analysis techniques**

*2.4.1 Nitrogen isotopic ratio in the $N_2$ initial gases*

Two $N_2$ tanks were used in the PAMPRE experiment: pure $N_2$ and a $N_2$-$CH_4$ mixture (90-10%), which were mixed up in the flow in order to get 1, 2, 5 and 10% $CH_4$ in the initial gas mixture. One tank of pure $N_2$ was used in the Nebulotron setup, which was mixed up with pure CO. The isotopic compositions of the two pure $N_2$ tanks (PAMPRE and Nebulotron) were obtained by filling a glass vessel with $N_2$ gases from the respective tanks and by analyzing it by dual-inlet technique on a MAT253 mass-spectrometer at CRPG (Nancy, France).

The $N_2$-$CH_4$ tank used in the PAMPRE experiment could not be directly measured because of its high $CH_4$ content. In order to remove $CH_4$ from the $N_2$, an aliquot of this mixture was oxidized during 45 min in a CuO furnace at 900°C to oxidize $CH_4$ into $CO_2$, which was then removed by adsorption on a cold trap held at -172°C. The CuO furnace was then cooled down to 450°C and the purified $N_2$ gas was transferred to a Cu furnace at 600°C in order to trap residual oxygen that could still be present in the line. The yield of the purification was close to 100% thus implying no isotope fractionation effect during purification. The purified $N_2$ was then measured by dual-inlet mass spectrometry for its isotopic composition.

Data were normalized against the NSVEC-air standard (International Atomic Energy Agency, Vienna, Austria). Blanks were negligible and errors (±0.04 ‰ on $\delta^{15}N$, 1σ) include external reproducibilities on the standards and on the samples obtained following 10 analyses of each gas.

*2.4.2 Bulk nitrogen content and isotopic ratio of aerosols by EA-IRMS*

Bulk measurements of nitrogen content and isotopic ratio of aerosols were performed with an elemental analyzer (EuroVector) coupled to an isotope ratio mass spectrometer (IsoPrime, GV Instruments) at CRPG (Nancy, France). $N_2$ (purity 99.999%) was used as the reference gas. The elemental analyzer was calibrated with urea (46.65% N), glutamic acid (9.52% N) and ammonium sulphate (21.2% N) for nitrogen in the range of 0.04-0.18 mg. Certified nitrogen isotopic standards (IAEA-N-1, $\delta^{15}N$ = 0.4 ± 0.2‰ and USGS-25, $\delta^{15}N$ = -30.4 ± 0.4‰, International Atomic Energy Agency, Vienna, Austria) were used to calibrate the mass spectrometer and correct sample isotopic ratios. Two aliquots of each aerosol sample was analyzed and bracketed by several standards (elemental and isotopic standards) analysis. The external reproducibility of isotopic standards was 0.2 ‰ (1σ) on the $\delta^{15}N$ values.

*2.4.3 Stepwise pyrolysis extraction and static mass spectrometry*

We used a static stepwise pyrolysis extraction and a static mass spectrometry method (Marty and Humbert 1997) in order to check the evolution of the nitrogen isotopic composition of aerosols during their thermal degradation. Aerosols were wrapped in Pt foil and pre-heated at 100°C for 24h under high vacuum (< $10^{-8}$ mbar) before extraction. Aerosols were thermally degraded by stepwise pyrolysis up to 950°C (3 to 5 steps, each of 20 min long) in a double-walled quartz tubing furnace (Yokochi et al., 2009). Extracted gases were purified by oxidation through a CuO furnace at 800°C. Most oxidized gases were removed using a cold trap held at -172°C, and only $NO_x$ compounds and minor amounts of CO and organic volatiles remained in the purification line. The CuO furnace was then cooled to 450°C to reduce $NO_x$ to $N_2$. Purified $N_2$ was then expanded and analyzed in a mass spectrometer working in static mode (Marty and Humbert, 1997; Marty and Zimmermann, 1999). $\delta^{15}N$ external reproducibility on air standards was 1.8‰ (1σ). Blanks were less than 1% of the samples. Two samples from the PAMPRE experiment (1% and 5% $CH_4$) and an aliquot of the Nebulotron aerosols were analyzed with this technique (Fig. 5).

## 3 Results

### 3.1 Elemental composition of aerosols: H/C, N/C and O/C ratios

Fig. 2 shows the H/C, N/C and O/C ratios of the Nebulotron aerosols, as well as the previously reported data by Sciamma-O'Brien et al., 2010 for the PAMPRE aerosols. Interestingly, the H/C values of the PAMPRE aerosols increase while their N/C values decrease with increasing $CH_4$ concentration in the initial gas mixture. The oxygen content of these aerosols is interpreted as resulting from water adsorption and/or from partial oxidation in air after recovery (Sciamma-O'Brien et al., 2010). In contrast to the PAMPRE aerosols, the Nebulotron aerosols are characterized by a high O/C ratio, consistently with the high initial concentration of CO. As no hydrogen was present in the $CO-N_2$ gas mixture of the Nebulotron experiment, the high H/C value of the Nebulotron aerosols may only come from water contamination. Indeed, as suggested before, the complete removing of the water adsorbed on the reactor walls would have required baking the reactor and secondary pumping.

Moreover, solids that are produced in plasmas are known to be highly reactive: water adsorption on the Nebulotron solids when put out to the atmosphere cannot be excluded. The Nebulotron and the PAMPRE aerosols exhibit similar N/C ratios despite an initial $N_2$ concentration in the Nebulotron experiment ~5 times lower than in the PAMPRE one. This highlights a nitrogen incorporation from gas to solid much more efficient in the Nebulotron setup than in the PAMPRE one, probably due to the higher energy to volume ratio in the former relative to the PAMPRE one (Table 1).

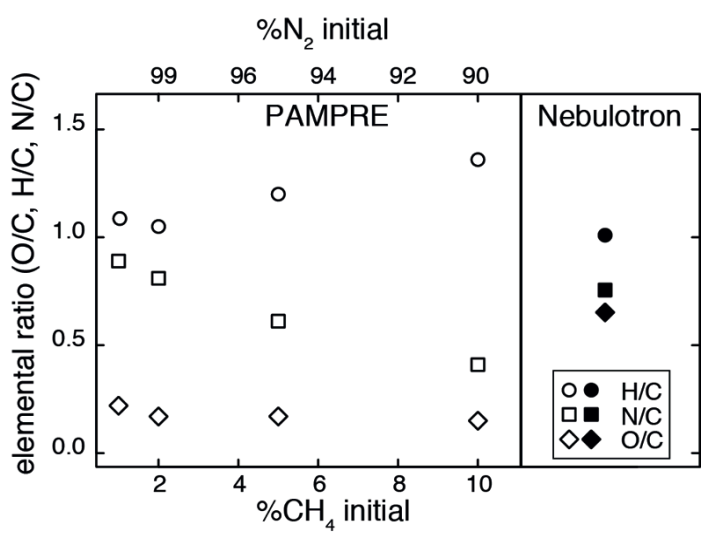

**Fig. 2**: Elemental composition of the PAMPRE (open symbols) and Nebulotron (filled symbols) experiments. Elemental ratios of the PAMPRE aerosols are presented as a function of the initial $CH_4$ concentration in the gas mixture (data from Sciamma-O'Brien et al., 2010). Error bars are smaller than the symbol sizes.

**3.2 STXM based XANES spectroscopy: carbon and nitrogen speciation**

Synchrotron-based XANES spectroscopy at the C and N K-edges provides key insights on carbon and nitrogen speciation in the PAMPRE and Nebulotron aerosols. These two aerosol mixtures are spectroscopically very homogeneous: over 20 particles of each aerosol, 19 exhibit very similar C- and N-XANES spectra. Representative C- and N-XANES spectra are shown in Fig. 3.

At the C K-edge, the low intensity of the absorption feature that can be seen in both samples at 285.1 eV (C1) is commonly attributed to electronic transitions of aromatic or olefinic carbon groups (C=C) (Bernard et al., 2010a, Bernard et al., 2010b). This is consistent with previous chemical studies which have pointed out an organic structure dominated by methyl groups and sp$^2$ carbons; and a lack or low abundance of any protonated aromatic or heteroaromatic rings in the PAMPRE aerosols (Quirico et al., 2008, Derenne et al., 2012). The peak at 286.8 eV (C2), identified in C-XANES spectra of both samples, is generally assigned to electronic transitions of ketonic or phenolic groups (C=O) and/or nitrile groups (C≡N) (Cody et al., 2008; Solomon et al., 2009). Both functions may be present within Nebulotron areosols, while only nitrile groups contribute to this peak in spectra of PAMPRE aerosols as attested by Infra-Red spectroscopy data (Quirico et al., 2008, Gautier et al., 2012). The peak at 288.1 eV (C3), also seen in spectra of both types of aerosols, is usually assigned to electronic transitions of amidyl groups (CO-NH$_x$) (Cody et al., 2008, Nuevo et al., 2011) and/or aliphatics (CH$_{1-3}$) (Bernard et al., 2012a, Bernard et al., 2012b; Buijnsters et al., 2012). In contrast to the sharp and intense peak that can be seen in the spectra of the Nebulotron aerosols, for which both of these functionalities may contribute, the broader absorption feature located at 288.1 eV in spectra of the PAMPRE aerosols can only be related to the absorption of various types of aliphatics, mostly connected to nitrogen atoms as indicated by FTIR and NMR data (Quirico et al., 2008, Gautier et al., 2012, Derenne et al., 2012). Finally, the peak located at 289.3 eV (C4), only observed in spectra of Nebulotron aerosols, is attributed to electronic transitions of hydroxylated- or ether-linked C species (Cody et al., 2008; Solomon et al., 2009), confirming the oxidized nature of these aerosols.

These results notably show that a part of the hydrogen incorporated into the Nebulotron aerosols is linked to carbon atoms in aliphatic and aromatic functions. This confirms the dissociation of a hydrogenated molecule, most probably H$_2$O, and further active chemistry to form C−H bonds into the plasma. Even if H$_2$O partial pressure in the plasma is small, hydrogen is probably very efficiently incorporated into the organic structure, for which hydrogen is necessary to build organic bonds and for polymerization. The presence of terminal −OH functions (C4 peak) in the Nebulotron aerosols' spectra likely traces oxidation and water adsorption on the aerosols' surface when they are extracted from the reactor.

At the N K-edge, three main peaks have been identified in the N-XANES spectra of the PAMPRE and Nebulotron aerosols at 398.1 eV (N1), 399.7 eV (N2) and ~ 401 eV (N3) (Mitra-Kirtley et al., 1993, Mullins et al., 1993, Leinweber et al., 2007; Cody et al., 2008; Nuevo et al., 2011). These peaks can be attributed to electron transitions of imine (C=N), nitrile (C≡N) and amidyl groups (CO-$NH_x$), respectively. Nitrogen-bearing heterocycles may also contribute to those peaks (Mitra-Kirtley et al., 1993, and references therein). Saturated amines have been identified in the PAMPRE aerosols in FTIR data (Quirico et al., 2008, Gautier et al., 2012). In XANES spectra, saturated amines generally appear at ~ 406 eV (Mitra-Kirtley et al., 1993, Mullins et al., 1993) and their contribution to the Nebulotron and the PAMPRE aerosols spectra is certainly masked by the diffuse absorption observed at the same energy likely corresponding to highly delocalized excited states or to the overlapping contribution of electronic and atomic resonances (Feshbach resonances, Stöhr, 1992).

Interestingly, although nitrogen in the Nebulotron aerosols appears to be mainly within amidyl groups whereas the PAMPRE aerosols are richer in imine and nitrile groups, nitrogen seems entirely linked to C or H atoms, as no N−O functionalities have been identified. This suggests that –NH and –CN bonds are favored despite the high oxygen content in the starting gas mixture of the Nebulotron. Thus, the volatile precursors of the nitrogen incorporation into the aerosols might be comparable in the Nebulotron and the PAMPRE experiments.

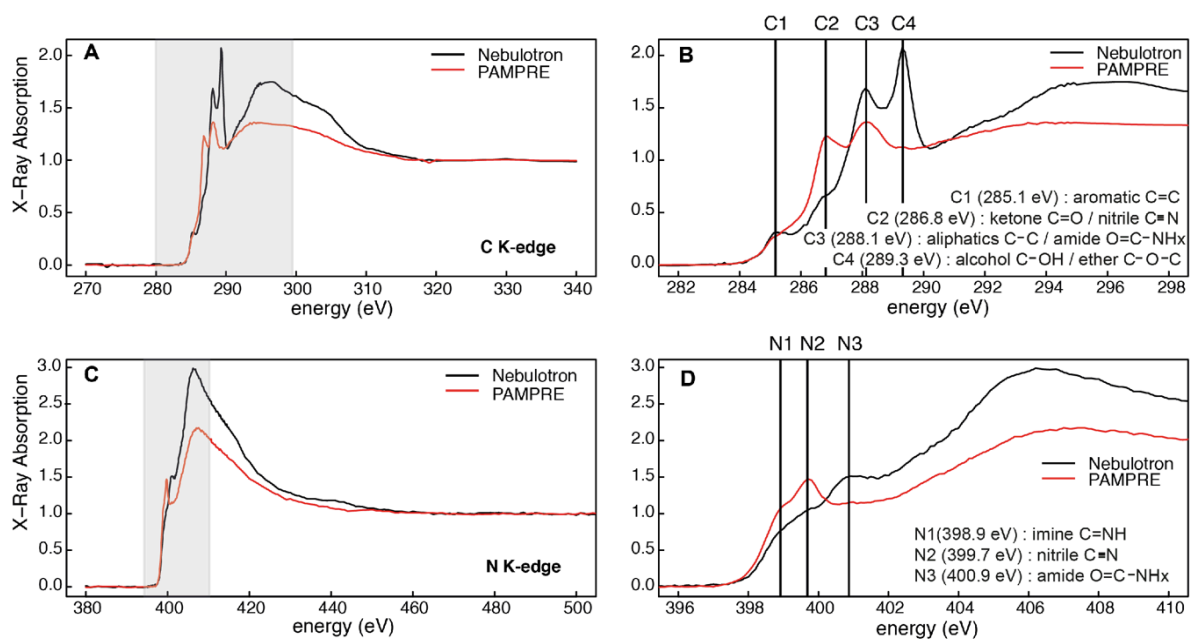

**Fig. 3**: XANES spectra and band assignment in the C- and N-K edges ((a) and (c) respectively) of the PAMPRE (5% initial CH$_4$) and Nebulotron aerosols. (b) and (d) are a zoom of the shaded area in the (a) and (c) plots respectively.

### 3.3 Nitrogen isotopic composition of aerosols

Because the nitrogen isotopic ratio of initial N$_2$ in the experimental gas mixtures is slightly different from that of air ($\delta^{15}$N=0‰ by definition), we use the $\Delta$ notation, which represents the difference between the isotopic composition of the products (aerosols) and that of the reactant (initial gas mixture):

$$\Delta^{15}\text{N} = \delta^{15}\text{N-aerosols} - \delta^{15}\text{N-initial gas (in ‰)}$$

The $\delta^{15}$N values measured by EA-IRMS for the experimental aerosols from the PAMPRE and the Nebulotron setups and the corresponding $\Delta^{15}$N values are reported in the Table 3.

| experimental setup | initial gas mixture | δ¹⁵N (‰) initial N₂[a] | σ | δ¹⁵N (‰) aerosols | σ | Δ¹⁵N (‰) gas-aerosols* | σ |
|---|---|---|---|---|---|---|---|
| PAMPRE | CH₄-N₂ (10-90) | -3.18 | 0.04 | -17.8 | 0.2 | -14.6 | 0.2 |
| PAMPRE | CH₄-N₂ (5-95) | -3.42 | 0.04 | -24.6 | 0.2 | -21.2 | 0.2 |
| PAMPRE | CH₄-N₂ (2-98) | -3.56 | 0.04 | -26.0 | 0.2 | -22.4 | 0.2 |
| PAMPRE | CH₄-N₂ (1-99) | -3.60 | 0.04 | -28.1 | 0.2 | -25.5 | 0.2 |
| Nebulotron | CO-N₂ (80-20) | -3.4 | 0.2 | -26.2 | 0.2 | -22.8 | 0.2 |

[a] The $\delta^{15}N$ of $N_2$ of the initial the gas mixture for the four PAMPRE experiments has been calculated by mass balance between the $\delta^{15}N$ of the two tanks which are -3.65 ± 0.04‰ and -3.18 ± 0.04‰ for the $N_2$ tank and the $N_2$-$CH_4$ tank respectively.

*$\Delta^{15}N$ gas-aerosols = $\delta^{15}N$aerosols - $\delta^{15}N$gas.

**Table 3**: Nitrogen isotopic composition in initial gas mixtures and aerosols from the PAMPRE and Nebulotron experiments from EA-IRMS measurements.

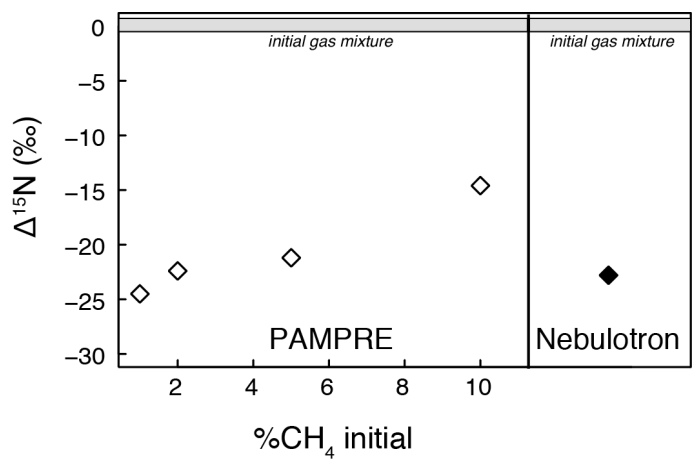

**Fig. 4**: Bulk nitrogen isotopic composition of the PAMPRE (open diamonds) and Nebulotron (filled diamonds) aerosols, expressed as a $\Delta^{15}N$ (in ‰) relative to the isotopic composition of the initial $N_2$ gas (represented by the shaded area). $\Delta^{15}N$ of the PAMPRE aerosols is presented as a function of the initial $CH_4$ concentration in the gas mixture. Error bars (1σ) are smaller than the symbol sizes.

All aerosols show a systematic depletion in the heavier nitrogen isotope $^{15}N$ compared to the initial gas $N_2$, with $\Delta^{15}N$ ranging from -15 to -25‰ whatever the setup used (Fig. 4). Previous $\Delta^{15}N$ values of -3 to -17‰ for nitrogen trapped in plasma-synthetized organics were reported by Kerridge et al., 1989. However, neither these data nor the experimental setup were discussed. Step-heating data (Table 4, Fig. 5) confirm such negative values and show overall a good homogeneity for N isotope fractionation, except for the lowest temperature step of the Nebulotron experiment (Fig. 5, top panel), which value is closer to air and may result from atmospheric nitrogen contamination.

Overall, isotopic data from step-heating and bulk measurements show a remarkable consistency, implying that these organic aerosols are globally isotopically homogeneous, whatever the setup production and the oxidation degree of the initial gas mixture.

| sample | mass (µg) | extraction temperature (°C) | N (moles) x$10^{-8}$ | σ | % of total N extracted | $\delta^{15}$N air (‰) | σ |
|---|---|---|---|---|---|---|---|
| PAMPRE (5% CH$_4$) (N$_2$-CH$_4$) | 66 | 150 | 0.024 | 0.001 | 0.07 | -24.5 | 2.8 |
| | | 420 | 1.40 | 0.04 | 4.0 | -23.8 | 2.4 |
| | | 650 | 4.30 | 0.12 | 12.2 | -20.5 | 2.0 |
| | | 900 | 19.5 | 0.5 | 55.4 | -26.9 | 2.1 |
| | | 950 | 10.0 | 0.3 | 28.4 | -20.0 | 2.2 |
| | | | | | | **-24.1*** | **2.1** |
| PAMPRE (1% CH$_4$) (N$_2$-CH$_4$) | 47 | 185 | 0.004 | $1.10^{-4}$ | 0.01 | *-16.9* † | *24.3* |
| | | 500 | 4.16 | 0.11 | 8.5 | -28.6 | 2.1 |
| | | 900 | 33.0 | 0.9 | 67.8 | -29.9 | 2.8 |
| | | 930 | 11.5 | 0.3 | 23.6 | -29.4 | 2.4 |
| | | | | | | **-29.7** | **2.7** |
| Nebulotron (CO-N$_2$) | 79 | 150 | 0.001 | $3.10^{-5}$ | 0.01 | *-1.4* | *47.4* |
| | | 420 | 0.59 | 0.02 | 3.6 | -18.9 | 2.5 |
| | | 650 | 10.8 | 0.3 | 65.7 | -23.7 | 2.2 |
| | | 800 | 1.52 | 0.04 | 9.2 | -25.5 | 2.0 |
| | | 930 | 3.53 | 0.1 | 21.5 | -27.3 | 2.1 |
| | | | | | | **-24.4** | **2.1** |

† Italic values: extracted nitrogen was too low and blank contribution exceeded 10% of the analysed nitrogen.

* Bold values: $\delta^{15}$N weighted average upon heating-steps

**Table 4**: Nitrogen isotopic composition in aerosols from the PAMPRE and Nebulotron experiments from step-heating and static mass-spectrometry measurements.

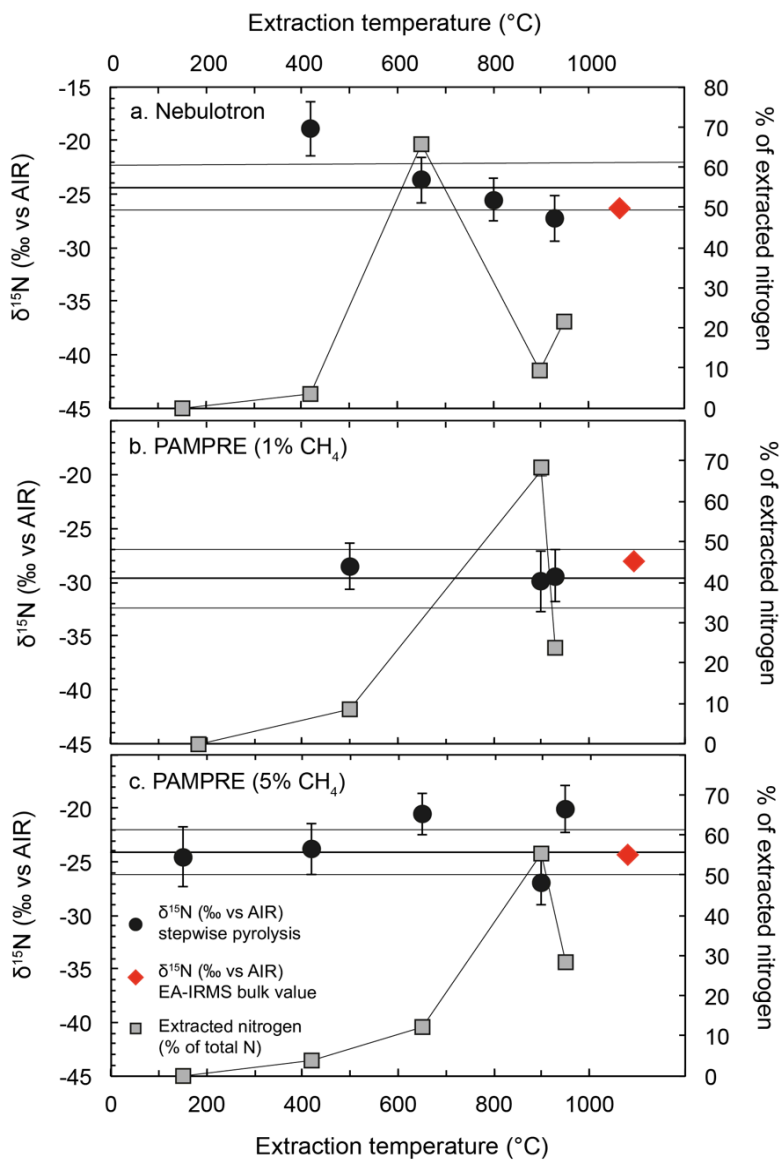

**Fig. 5**: Nitrogen isotopic composition ($\delta^{15}$N in ‰, left axis, *dark circles*) and quantity of released nitrogen (in % of total nitrogen extracted, right axis, *grey squares*) as a function of pyrolysis temperature of the Nebulotron and the PAMPRE aerosols. *Solid lines*: weighted average and standard deviation of the nitrogen isotopic ratios over all extraction steps. *Red diamonds*: EA-IRMS bulk $\delta^{15}$N measurements (error bars (1σ) are smaller than the symbol size).

# 4 Discussion and implications

## 4.1 Origin of the nitrogen isotopic fractionation in plasma experiments

The PAMPRE and the Nebulotron aerosols share comparable $^{15}$N depletions relative to the initial N$_2$ gas around -20‰, although experimental setups have different geometry, carbon source and electric discharge characteristics. This heavy isotope depletion is relatively large compared to the natural nitrogen isotopic variations observed in terrestrial organic molecules or geological objects, for which $\delta^{15}$N lower than -10‰ are rare (Coplen et al., 2002). No such $^{15}$N-depleted values were reported for abiotic and open-system experiments comparable to the PAMPRE and Nebulotron simulations performed in the present study.

However, the experimental nitrogen isotopic fractionation measured in this study is small compared to the large isotopic variations that are observed in planetary atmospheres, e.g., Titan, or among primitive reservoirs and objects of the solar system, To explain such variations, several studies have focused on the dissociation of $N_2$ by photons in the far UV range (Liang et al., 2007; Croteau et al., 2011; Lyons et al., 2009; Chakraborty et al., 2013). Photodissociation of $N_2$ could lead in principle to large $^{15}$N enrichments (several 100 to several 1000 ‰) either by (i) isotopic selective photoabsorption and further self-shielding of $N_2$ in a dense atmosphere (Liang et al., 2007, Lyons et al., 2009) and/or by (ii) indirect (also called accidental) predissociation of $N_2$ excited states (Lorquet and Lorquet, 1974, Lefebvre-Brion and Field, 2004, van de Runstraat et al., 1974, Muskatel et al., 2011). In contrast, in the present study, (i) the N isotope fractionation leads to compounds depleted, and not enriched, in $^{15}$N, and (ii) the isotopic effect is much more modest than expected to account for N isotope variations in the Solar System.

In plasmas, $N_2$ dissociation mostly occurs via (i) electron impact at electron energies above the dissociation threshold of $N_2$ (9.8 eV) and (ii) heavy particles collisions (Fridman, 2008, Alves et al., 2012). As electron energy of impacting electrons is continuous and not quantified as for photons, several excited states of atoms and molecules can be reached by electron impact in plasmas. As a result, electron impact cross-sections of $N_2$ dissociation are not as structured as photoabsorption cross-sections (Zipf and McLaughlin, 1978). Because of the continuous distribution of electron energies, $^{14}N_2$ and $^{14}N^{15}N$ dissociation rates should not be different. Moreover, atoms and molecules in plasma are not subjected to self-shielding, as electrons are isotropically produced by the electric discharge, in contrast to what occur in photon-induced reactions.

Although indirect predissociation of $N_2$ might be invoked, this scenario has never been documented in plasma electron energy ranges (0-20 eV) relevant to the present study. In any case, this dissociation mechanism is quite rare in plasma, and isolated events may certainly not affect the final nitrogen isotopic composition of aerosols, even with an isotopic fractionation of a few thousands of ‰ (Muskatel et al., 2011).

Hence, other processes are responsible for the $^{15}N$-depletion observed in the Nebulotron and PAMPRE aerosols. These $^{15}N$-depleted values may result from one or several fractionation processes all along the reaction chain that leads to aerosol formation favoring the light $^{14}N$ isotope to be produced in the plasma by $N_2$ dissociation and/or isotopically light molecules to be incorporated into the aerosols via polymerization of gaseous precursors. Because the two setups and both type of aerosols are very different (e.g., elemental abundances and XANES characteristics) the fractionation process(es) must be independent of the chemistry and the oxidation degree of the environment.

Kinetic Isotopic Fractionation (KIF, O'Neil, 1986, Li et al., 2009) is likely to occur during such experiments, as the plasmas used in the present study are out of equilibrium. Chemical reactions occurring in the gas phase and during aerosol polymerization are largely controlled by kinetic competitions (Fridman, 2008). KIF during aerosol polymerization was already invoked to account for the $^{13}C$ isotopic fractionation observed in plasma-synthetized carbonaceous solids (Des Marais et al., 1981, Chang et al., 1983).

KIF is a mass-dependent fractionation process. Considering two molecules, *A-$^{14}$N* and *B-$^{15}$N*, isotopically substituted, and assuming, to a first approximation, a Maxwell-Boltzmann distribution of velocities in the gas phase in our plasma setups, the KIF α factor can be written as:

$$\alpha = \frac{k_B}{k_A} \approx \frac{<v_B>}{<v_A>}$$

with *k* being the reaction coefficient and *v* the average relative velocity within the gas.

The molecular velocity *v* will be proportional to $T^{1/2}$ and $m^{-1/2}$, with *T* and *m* being the gas temperature and the mass of the molecule, respectively. For an isothermal gas, α is then equal to :

$$\alpha = \frac{k_B}{k_A} \approx \sqrt{\frac{m_A}{m_B}}$$

An ε value (in ‰) commonly defines the isotopic enhancement (or depletion depending on the value of α relative to 1) related to an isotopic fractionation factor α as:

$$\varepsilon = (\alpha - 1) \times 1000$$

ε is the theoretical equivalent of our experimental $\Delta^{15}N$ mentioned above.

As a result, the $^{14}N$-substituted molecules will react faster than the $^{15}N$-substituted ones. Furthermore, the isotopic fractionation ε will be larger if low mass molecules are reacting.

The Nebulotron and the PAMPRE aerosols have N isotopic fractionations between -15 and -25‰ relative to initial $N_2$. Considering a KIF occurring during the aerosols' growth and using the equations above, such a $^{15}N$-depletion is compatible with involved N-bearing molecules with masses ranging theoretically between 19 and 33 amu. HCN (m = 27, ε = -18‰) and $CH_2NH$ (m = 29, ε = -17‰), two molecules identified in the PAMPRE gas phase (Gautier et al., 2011, Carrasco et al., 2012), can contribute to the formation of C≡N, C=NH and C−$NH_2$ bonds identified in both the PAMPRE and the Nebulotron aerosols. These two molecules seem to be the main precursors for the nitrogen incorporation into the aerosols in the two setups. Except for the 10% $CH_4$-PAMPRE aerosol ($\Delta^{15}N \sim -15‰$), all the analyzed aerosols exhibit $\Delta^{15}N$ lower than -21‰ in both the PAMPRE and the Nebulotron setups. In such an "aerosol growth isotopic fractionation" model, this requires the involvement of molecules than can yield a larger $^{15}N$-depletion than the one theoretically provided by HCN and $CH_2NH$. This can be provided by smaller molecules like $NH_3$ (m = 17, ε = -28‰) or even atomic N (m = 14, ε = -34‰), two species that are highly reactive too and that were identified in the PAMPRE experiment as well (Alves et al., 2012; Carrasco et al., 2012). On the other hand, the 10% $CH_4$-PAMPRE aerosols at -14.6‰ require a heavier molecule than HCN (or $CH_2NH$) that could imprint a less $^{15}N$-depleted signature. In the PAMPRE setup context, one could think of acetonitrile ($CH_3CN$, m = 41, ε = -12‰) or even larger N-bearing molecules with 4, 5 or more carbon atoms that were identified in the gas phase of the PAMPRE experiment (Gautier et al., 2011, Carrasco et al., 2012).

We thus suggest that the $^{15}N$-depleted isotopic composition of the experimental aerosols relative to the initial $N_2$ is due to a KIF during the aerosol growth by polymerization of N-bearing molecules, mainly HCN and/or $CH_2NH$. The strong relationship that is observed between the PAMPRE aerosols nitrogen isotopic composition and the initial $CH_4$ concentration at which they were produced (Fig. 4) requires the involvement of other N-bearing molecules that will enhance (at low $CH_4$ %) or reduce (at high $CH_4$ %) the $^{15}N$-depleted signature of HCN and $CH_2CN$; namely $NH_3$ (or N) and $CH_3CN$ (or even larger N-bearing molecules), respectively. Such an hypothesis is in good agreement with FTIR and High Resolution Mass Spectrometry (HRMS) data obtained on the PAMPRE aerosols (Gautier et al., 2012; Gautier, 2013), which clearly show an increase of the saturated amine (C−$NH_2$ bonds) content and an increase of light N-rich compounds incorporated into the aerosols as the initial $CH_4$ % decreases.

Isotopic fractionation during synthesis of organic aerosols in plasma is thus consistent with kinetic isotopic fractionation (KIF). KIF is not specific to plasma and is likely to occur in every abiotic system, whether it is an electron, photon or temperature dominated environment. However, larger isotope effects in specific conditions such as photodissociation in the far UV light range may hide its $^{15}$N-depleted typical signature.

## 4.2 Cosmochemical and planetary implications

- *Nitrogen in meteoritic organics*

Most of the organic compounds encountered in primitive meteorites are insoluble poly-aromatic macromolecules, referred as the so-called "Insoluble Organic Matter" (IOM, Derenne and Robert, 2010). Bulk nitrogen isotopic compositions of these organics show little variations and are very close to the terrestrial value of $\delta^{15}N = 0‰$, although some rare compounds exhibit extreme isotopic heterogeneities at very fine scales ("hot-spots", Busemann et al., 2006, Briani et al., 2009, Bonal et al., 2010). The process, timing and location of the formation of these large isotopic fractionations are still not resolved, but are probably closely related to the processes having led to the formation of these meteoritic organic compounds, processes that are still poorly understood (Alexander et al., 2007, Bonal et al., 2010, Hily-Blant et al., 2013).

The Nebulotron organic aerosols were experimentally produced from a gas mixture relatively close to the protosolar gases ($N_2$-CO *with $H_2O$ contamination*). As discussed above, the use of an electric discharge to dissociate $N_2$ does not allow the production of very large isotopic fractionations as expected for UV photons. Thus, the reactions occurring in the Nebulotron setup do not explain the extreme $^{15}$N-enrichments observed at the micron scale in the IOM nor the bulk nitrogen isotopic composition of these organics (enriched by ~600 ‰ relative to the protosolar nebula composition). In this case, processes such as $N_2$ photodissociation/self-shielding or low temperature ion-molecule reactions seem more suitable to address these large $^{15}$N-enrichments.

- *Titan's aerosols*

On Titan, the dissociation in the ionosphere of the two main gases $N_2$ and $CH_4$ results in an efficient integration of nitrogen in the gas products, as shown by the detection of numerous N-containing neutrals and ions by instruments on board Cassini, both on the dayside and the night side (Cui et al., 2009, Nixon et al., 2013). Aerosols are initiated in the upper atmosphere, where a dusty plasma chemistry occurs (Lavvas et al., 2013) possibly supporting the significant nitrogen amount detected in Titan's aerosols close to the surface by the Huygens probe (Israël et al., 2005). The PAMPRE experiment seems well representative of the chemistry occurring in Titan's atmosphere, as both gas and solid products are consistent with Cassini data (Gautier et al., 2011, Carrasco et al., 2012, Carrasco et al., 2013).

Although the dominant energy source available for $N_2$ dissociation in the upper Titan's atmosphere is solar radiation, suprathermal electrons may play a significant role in the neutral and ion chemistry occurring in Titan's atmosphere. Indeed, electron impact may account for ~10% of the total $N_2$ dissociation rate during dayside, at an altitude of ~1000 km, where initiation of aerosol growth is expected (Lavvas et al., 2011). On the nightside, solar photons cannot operate and electrons are the main source for ionization and dissociation of atmospheric molecules, despite a lower density than for the dayside (Cravens et al., 2009, Ågren et al., 2009).

Based on the present experimental KIF data, isotopic effects in the non-photonic dissociation of $N_2$ and further aerosol synthesis are not expected to play a major role in determining the large $^{15}N$-enrichment of HCN relative to $N_2$ ($\Delta^{15}N \sim 4200‰$) in Titan's upper atmosphere (Vinatier et al., 2007, Liang et al., 2007; Niemann et al., 2010). However, the PAMPRE nitrogen isotopic data may give new constraints for nitrogen isotopic modeling in Titan's atmosphere, knowing that nitrogen isotopic composition of Titan's aerosols has not been measured yet. $N_2$ photodissociation and self-shielding isotopic effects calculated by Liang et al., 2007 resulted in a factor of ~ 2.4 too large $^{15}N/^{14}N$ ratio for HCN ($^{15}N/^{14}N$ = 4.35 $10^{-2}$, Liang et al. 2007) compared to the observations ($^{15}N/^{14}N$ = 1.79 ± 0.26 $10^{-2}$, Vinatier et al. 2007). These authors thus suggested that an additional source of unfractionated atomic nitrogen, produced by ion/electron-impact induced $N_2$ dissociation was required to counterbalance the too large effect of self-shielding. Isotopic effects described in the present study may contribute to counterbalance the calculated self-shielding $^{15}N$-enrichment. We can roughly quantify by a simple mass balance calculation the importance of KIF for nitrogen isotopes in Titan's atmosphere. Assuming a -20‰ isotopic kinetic effect with 10 % of electron-impact dissociated $N_2$ on the dayside (Lavvas et al. 2011) and 100% of electron-impact dissociated $N_2$ on the nightside, such a calculation provides a final $^{15}N/^{14}N$ ratio of 2.25 $10^{-2}$, dividing by 2 the discrepancy between the HCN isotopic ratio calculated by the self-shielding model and the measured one. This mass-balance suggests that KIF may play a role in the nitrogen isotopes balance in Titan's atmosphere. Besides, as first pointed out by Croteau et al., 2011, photoionization, photodissociation and non-photonic dissociation of $N_2$ provide distinct nitrogen isotopic fractionation signatures. The results of the present work may thus allow discriminating between putative aerosol formation pathways when nitrogen isotopic measurements in Titan's aerosols and organic molecules will become feasible.

- *Organics on early Earth*

The Earth's Archaean atmosphere (3.8 to 2.5 Ga ago) was anoxic and mainly consisted of $N_2$ and $CO_2$, with possibly minor amounts of $H_2$ and $CH_4$ (Zahnle, 1986, Pavlov et al., 2001, Tian et al., 2005, Wordsworth and Pierrehumbert, 2013, Charnay et al., 2013; Marty et al., 2013). In the presence of methane, a Titan-like hydrocarbon haze might have existed before widespread oxygenation around 2.5 Ga ago (Zahnle, 1986, Sagan and Chyba, 1997, Domagal-Goldman et al., 2008, Tian et al., 2011, Zerkle et al., 2012,). Organic aerosol formation in such mildly reduced atmospheres has been demonstrated experimentally (Trainer et al., 2006,

DeWitt et al., 2009). The formation of organic aerosols may also account for the isotopic fractionation of xenon in the ancient atmosphere (Hébrard and Marty, 2014) evidenced from the analysis of ancient sedimentary rocks (Pujol et al., 2011).

Without an ozone layer, the EUV photons from the Sun might have triggered atmospheric photochemical reactions such $N_2$ photodissociation and organic production. Even though the present experimental atmospheres ($N_2$-$CH_4$ and $N_2$-CO-$H_2O$) do not represent early Earth's atmosphere composition, it seems reasonable to assume that, in the expected mildly oxidized early Earth's atmosphere, active incorporation of nitrogen into organic aerosols have occurred as the result of photochemistry or lightning. The isotopic composition of Earth's atmospheric $N_2$ has not changed since the Archaean era, 3.5-3 Ga ago (Marty et al., 2013), confirming that the thin Earth's atmosphere has not been prone to nitrogen isotopic self-shielding as it is expected for Titan (Liang et al., 2007, Croteau et al., 2011). Thus, the present experimental results suggest strongly that, in the absence of such dramatic isotopic enrichments, nitrogen fixation into abiotic Archaean haze from atmospheric $N_2$ would have acquired a $^{15}N$-depletion down to -25‰.

In the Archaean, the atmospheric production of abiotic organic matter and its delivery to oceans and continental areas might have been low, but comparable to that produced by early biological activity (Kasting and Catling, 2003). Later, on the Neoarchaean Earth, abiotic organic haze might have co-existed with life and biogenic organic production (Zerkle et al., 2012). $^{15}N$-depleted abiotically formed aerosols could directly account for a significant fraction of sedimentary organics and thus leave their isotopic imprint in Archaean geological records. In line with this possibility, negative $\delta^{15}N$ values in the range of -11 to -6.2 ‰ measured in Paleoarchaean kerogens and metasedimentary rocks have been attributed to nitrogen cycle operated by microorganisms in a reducing environment (Beaumont and Robert, 1999, Pinti et al., 2001, Fig. 6). Alternatively, the present experimental results suggest that these low δ¹⁵N might be in part of abiotic origin. The progressive oxygenation of the Earth's atmosphere, leading to a haze-free atmosphere, resulted in a oceanic nitrogen cycle dominated by the biomediated denitrification process, and δ¹⁵N of organic matter evolved toward positive values (Boyd, 2001, Fig. 6). Considering the very negative δ¹⁵N of the abiotic organics synthetized in the present study, the gradual enrichment in ¹⁵N of organic matter from the Archaean to Present might also be the result of a transition from abiotic production to biotic one, and may therefore trace the temporal development of the biosphere.

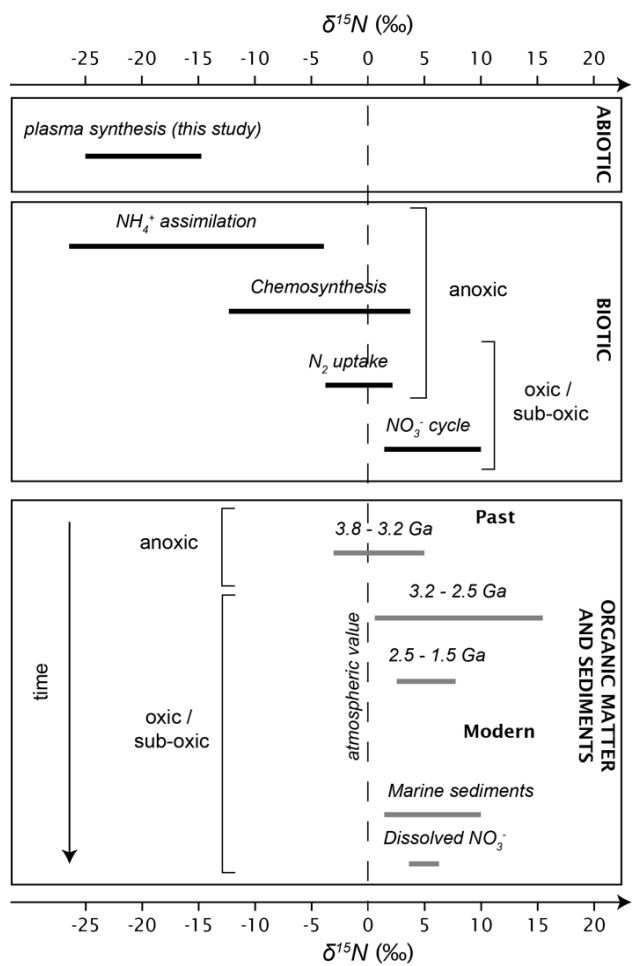

**Fig. 6**: Variations in the nitrogen isotopic composition of organic matter and biomass resulting from abiotic (this study, top box) and the main biomediated (middle box) processes. For comparison, δ¹⁵N measured in old kerogen and modern marine sediments are given (bottom box). Sources of the data: $NH_4^+$ assimilation (Hoch et al., 1992, Macko et al., 1987, Waser et al., 1998), chemosynthesis (Brooks et al., 1987, Conway et al., 1994), N₂ uptake (Macko et al., 1987, Zerkle et al., 2008), $NO_3^-$ cycle (Macko et al., 1987, Mariotti et al., 1981, Boyd, 2001), Past kerogen data (compilation from Thomazo and Papineau, 2013, average

$\delta^{15}N\pm\sigma$), Modern marine sediments and dissolved $NO_3^-$ (Peters et al., 1978, Sweeney et al., 1978).

# 5 Conclusions

We have analyzed the nitrogen content, the N speciation and the N isotopic composition in organic aerosols synthetized by plasma from two starting gas mixtures containing $N_2$: $N_2$-$CH_4$ (PAMPRE setup) and $N_2$-CO (Nebulotron setup). This study demonstrates the efficient incorporation of nitrogen into the synthesized solids, independently of the oxidation degree and the $N_2$ content of the starting gas mixture. Both the PAMPRE and the Nebulotron aerosols show a comparable $^{15}N$ depletion of about -20‰ relative to the initial $N_2$. The experimental organic solids were found to be isotopically homogeneous, with negligible air contamination. This isotopic signature is attributed to mass-dependent kinetic isotopic fractionation, which could occur at any step in the synthesis of aerosols but most likely during polymerization. Such kinetic isotopic fractionation is too small and goes in the wrong direction to account for the large $^{15}N$ enrichments observed in some primitive Solar System materials compared to the protosolar $^{15}N/^{14}N$ value. Processes such as specific isotopic effects associated to $N_2$ photodissociation, self-shielding, or isotopic exchange at very low temperature remain the best candidates. Nonetheless, in environments where these processes do not occur, as in Titan's atmosphere on the nightside or in early Earth's atmosphere, kinetic isotopic effects may imprint a $^{15}N$-depleted signature. Such negative $\delta^{15}N$ values may characterize abiotic production of organics during the Archaean eon, rather than specific metabolic pathways.

**Acknowledgments**:

We thank L. Zimmermann (CRPG, Nancy) and V. Busigny (IPGP) for their help with the nitrogen isotopes measurements by stepwise pyrolysis and the $N_2$-$CH_4$ purification, respectively. C. France-Lanord kindly put at disposition the CRPG stable isotope facility. We are grateful to P. Cartigny for helpful discussions and to G. Cernogora for the "plasma lessons" and for his wise scientific comments. M.K. gratefully acknowledges support from the French Ministry of Higher Education through a PhD grant. This study was partly funded by the French Programme National de Planétologie (PNP) and by the European Research Council under the European Community's Seventh Framework Program (FP/7 2007-2013, grant agreement number 267255 to B.M.).

**Table captions:**

**Table 1**: Qualitative comparison of the two plasma setups used in this study.

**Table 2**: Experimental conditions of aerosol production in the PAMPRE and Nebulotron setups.

**Table 3**: Nitrogen isotopic composition in initial gas mixtures and aerosols from the PAMPRE and Nebulotron experiments from EA-IRMS measurements.

**Table 4**: Nitrogen isotopic composition in aerosols from the PAMPRE and Nebulotron experiments from step-heating and static mass-spectrometry measurements.

**Figure captions:**

**Fig. 1**: Schematics of the PAMPRE (a) and the Nebulotron (b) experimental setups.

**Fig. 2**: Elemental composition of the PAMPRE (open symbols) and Nebulotron (filled symbols) experiments. Elemental ratios of the PAMPRE aerosols are presented as a function of the initial $CH_4$ concentration in the gas mixture (data from Sciamma-O'Brien et al., 2010). Error bars are smaller than the symbol sizes.

**Fig. 3**: XANES spectra and band assignment in the C- and N-K edges ((a) and (c) respectively) of the PAMPRE (5% initial $CH_4$) and Nebulotron aerosols. (b) and (d) are a zoom of the shaded area in the (a) and (c) plots respectively.

**Fig. 4**: Bulk nitrogen isotopic composition of the PAMPRE (open diamonds) and Nebulotron (filled diamonds) aerosols, expressed as a $\Delta^{15}N$ (in ‰) relative to the isotopic composition of the initial $N_2$ gas (represented by the shaded area). $\Delta^{15}N$ of the PAMPRE aerosols is presented

as a function of the initial CH$_4$ concentration in the gas mixture. Error bars (1σ) are smaller than the symbol sizes.

**Fig. 5**: Nitrogen isotopic composition (δ$^{15}$N in ‰, left axis, *dark circles*) and quantity of released nitrogen (in % of total nitrogen extracted, right axis, *grey squares*) as a function of pyrolysis temperature of the Nebulotron and the PAMPRE aerosols. *Solid lines*: weighted average and standard deviation of the nitrogen isotopic ratios over all extraction steps. *Red diamonds*: EA-IRMS bulk δ$^{15}$N measurements (error bars (1σ) are smaller than the symbol size).

**Fig. 6**: Variations in the nitrogen isotopic composition of organic matter and biomass resulting from abiotic (this study, top box) and the main biomediated (middle box) processes. For comparison, δ$^{15}$N measured in old kerogen and modern marine sediments are given (bottom box). Sources of the data: NH$_4^+$ assimilation (Hoch et al., 1992, Macko et al., 1987, Waser et al., 1998), chemosynthesis (Brooks et al., 1987, Conway et al., 1994), N$_2$ uptake (Macko et al., 1987, Zerkle et al., 2008), NO$_3^-$ cycle (Macko et al., 1987, Mariotti et al., 1981, Boyd, 2001), Past kerogen data (compilation from Thomazo and Papineau, 2013, average δ$^{15}$N±σ), Modern marine sediments and dissolved NO$_3^-$ (Peters et al., 1978, Sweeney et al., 1978).